\documentstyle[12pt,a4]{article}

\def\hepawk/{\.{hepawk}}
\def\f77/{\.{FORTRAN-77}}
\def\hepevt/{\.{/hepevt/}}
\def\KRONOS/{\.{KRONOS}}
\def\KROWIG/{\.{KROWIG}}
\def\HERWIG/{\.{HERWIG}}

\def\KWVersion/{1.0}
\def\KRVersion/{1.2}
\def\HWVersion/{5.4}
\def\Date/{June 1991}

\newenvironment{example}[2]{
    \def\examplecaption{#1}
    \def\examplelabel{#2}
    \begin{figure}
    \begin{tabbing}
    \tab\tab\tab\tab\tab\tab\tab\tab\tab\tab\kill
  }{
    \end{tabbing}
    \caption{\examplecaption}
    \label{\examplelabel}
    \end{figure}
  }

\def\.#1{{\tt#1}}         
\def\,#1{{\rm\it#1\/}}
\def\C{\`\it}             
\def\tab{\quad\=}

\def\BS/{$\backslash$}

\thicklines

\title{\KROWIG/, Version \KWVersion/: \\
       Interfacing \KRONOS/ and \HERWIG/ \\
       or \\
       Higher Order Electromagnetic \\
       Radiative Corrections at HERA \\
       with Hadronic Final States
}

\author{{\sc Thorsten Ohl}%
                \thanks{Supported by
                a grant of Deutsche Forschungsgemeinschaft.}
                \thanks{preferred e-mail address: {\tt
                <ohl@crunch.ikp.physik.th-darmstadt.de>}}%
                \thanks{Address after Sept.~1, 1992: Lyman
                Laboratory of Physics, Harvard University,
                Cambridge, MA 02138, USA.}\\
      \hfil\\
      Deutsches Elektronen-Synchrotron DESY \\
      W-2000 Hamburg, Federal Republic of Germany
}

\date{}

\begin{document}
\maketitle

\vfill
\begin{abstract}
This manual describes version \KWVersion/ of the Monte Carlo
event generator \KROWIG/ for deep inelastic lepton hadron scattering
at HERA. \KROWIG/ combines the implementation of QED radiative
corrections in \KRONOS/ with the QCD parton showers and cluster
fragmentation of \HERWIG/.
\end{abstract}

\vbox to 0pt{\vss
  \vbox to \textheight{\noindent
    DESY 92-097 \hfill ISSN 0418-9833\\
    July 1992 \hfill
    \vfill
  }
}


\newpage
\section*{Program Summary:}

\begin{itemize}
\item{} {\bf Program name:} \KROWIG/.
\item{} {\bf Version:} \KWVersion/ (\Date/).
\item{} {\bf Author:} Thorsten Ohl, {\tt
    <ohl@crunch.ikp.physik.th-darmstadt.de>}
\item{} {\bf Long write-up:} This document.
\item{} {\bf Programming language:} \f77/.
\item{} {\bf Computer/operating system:}
  Any with a \f77/ environment.
\item{} {\bf Number of program lines:} $\approx 3000$
\item{} {\bf Other programs used:} \KRONOS/\cite{ADM+92a}
  (Version 1.2 or later),  \HERWIG/\cite{MW88} (Version 5.4),
  \.{PAKPDF}\cite{Cha92}.
\item{} {\bf Input files needed:} None.
\item{} {\bf Initial parton shower:} \HERWIG/.
\item{} {\bf Hard subprocesses generated:} Neutral current deep
  inelastic scattering.
\item{} {\bf Final parton shower:} \HERWIG/.
\item{} {\bf Fragmentation model:} Cluster (\HERWIG/).
\item{} {\bf Initial QED radiation:} Leptonic: leading logarithmic
  approximation, summed to all orders (\KRONOS/); quarkonic: N/A.
\item{} {\bf Final QED radiation:} Leptonic: leading logarithmic
  approximation, summed to all orders (\KRONOS/); quarkonic:
  included in the \HERWIG/ parton shower (optional).
\end{itemize}


\newpage
\section{Introduction}
\label{sec:intro}

The new $ep$ collider HERA will explore a new kinematical range for
deep inelastic lepton-hadron scattering (DIS) \cite{BI92}.
Due to the smaller mass of the electron and the higher available energy,
radiative corrections will be much larger than at the previous $\mu p$
experiments.  Inclusively, the leading contribution in each order of
perturbation theory can be estimated as
\begin{equation}
  \frac{\alpha_{QED}}{\pi} \ln\left(\frac{Q^2}{m^2_e}\right)
  \label{eq:LL}
\end{equation}
and the exclusive corrections can surpass 100\% in some regions of
phase space \cite{SAA+92}.
A quantitative understanding of the radiative corrections is therefore
mandatory for the physical interpretation of HERA data.
The electroweak corrections to deep inelastic $ep$ scattering have been
calculated completely to one loop order \cite{BS87,BBCR89} and the
leading logarithms of the next order have been given \cite{KMS91}.

However, for detailed experimental studies, the implementation of these
calculations in a Monte Carlo event generator is indispensable.  The
order $\alpha_{QED}$ corrections have been implemented in the {\tt
HERACLES} event generator \cite{KMS91a}, which has been interfaced in
{\tt DJANGO}\cite{SS92} to the {\tt LEPTO} QCD Monte
Carlo \cite{Ing86}.  At the parton level, the leading higher
order QED corrections have been implemented in the
\KRONOS/ Monte Carlo \cite{ADM+92a}.

\KROWIG/ extends the hadronically inclusive predictions of \KRONOS/ to
exclusive hadronic final states.  In addition, \KROWIG/ allows to
incorporate radiative corrections into the predictions of the \HERWIG/
model for QCD parton showers and cluster fragmentation
\cite{MW88}.  Thus \KROWIG/ is unique in two different aspects of HERA
physics:  it is the first Monte Carlo Event generator implementing QED
radiative corrections to DIS in the \HERWIG/ framework and it is also
the first Monte Carlo event generator implementing higher order QED
radiative corrections for hadronically exclusive final states.
Although multiphoton events will not play an important r\^ole during
the first years of experimentation at HERA \cite{ADM+HERA}, the
included soft photon exponentiation can have numerically appreciable
effects.

After elaborating on the physics of radiative corrections to exclusive
hadronic final states in section~\ref{sec:physics}, we shall discuss
their implementation in \KROWIG/ in section~\ref{sec:programming}.
The various parameters controlling \KROWIG/ are discussed in section
\ref{sec:parameters}.  Section~\ref{sec:f77} is devoted to the \f77/
interface for application programs.  Before concluding, we list in
section~\ref{sec:limitations} the limitations of \KROWIG/
version~\KWVersion/.
Technical details of the installation of \KROWIG/ and a sample
application are relegated in the appendix.

In order to avoid unnecessary duplication, this manual frequently
refers to the manuals for \KRONOS/ \cite{ADM+92c} and \HERWIG/
\cite{MWA+92} and assumes familiarity with the two programs.


\section{Implementation of \KROWIG/ \.{\KWVersion/}}
\label{sec:implementation}


\subsection{Physics issues}
\label{sec:physics}

Conceptionally, the interfacing of leptonic QED radiative corrections
and QCD Monte Carlos is simple, provided one stays in the framework of
the leading logarithmic approximation.  In this case the radiative
corrections factorize and can be presented to the QCD Monte Carlo as
{\em just another hard subprocess\/}.

In a nutshell, the reasoning is that the hadronically inclusive cross
section including radiative corrections is given by the parton model,
while QCD parton showers and
the subsequent fragmentation must happen with unit probability.

Factorization is in our approach equivalent to staying within the
usual one-photon-approximation, on which the QCD description of DIS
in terms of the improved parton model or operator product expansion is
built.  In particular this means that we will neglect the non-leading
box diagrams in which a second photon (or $Z^0$) is exchanged between
the leptonic and hadronic subsystems.

Using this approximation, the hadronically inclusive cross section can
be written as
\begin{eqnarray}
  \label{eq:x-sect}
  \frac{d^3\sigma}{dxdydz} (x,y,z)
    & = & \frac{\pi\alpha^2_{QED}}{x^{\prime2}y^{\prime2}s^{\prime2}}
      \sum_{i=q,\bar q} x' f_i(x',Q^{\prime2}) (A_i + (1 - y') B_i) \\
    &   & \qquad\qquad\qquad \times
      D(z,Q^2) \frac{\partial(x',y')}{\partial(x,y)} \nonumber
\end{eqnarray}
where
\begin{equation}
  x' = \frac{zxy}{y + z - 1}, \qquad
  y' = \frac{y + z - 1}{z}, \qquad
  s' = zs, \qquad Q^{\prime2} = z Q^2
  \label{eq:primed}
\end{equation}
In these formulae $x$ and $y$ are the (leptonic) Bjorken variables and
$z$ is the energy fraction of the incoming electron after initial
state radiation.  Final state radiation has been ignored here for
simplicity, but is included in \KROWIG/.  $A_i$ and $B_i$ denote the
appropriate combination of electroweak couplings and propagator
factors for the quark $i$.  $D$ and the $f_i$ are solutions of the
familiar evolution equations
\begin{eqnarray}
  \frac{\partial f_i(\xi,Q^2)}{\partial\ln Q^2}
    & = & \int\limits_\xi^1 \frac{d\eta}{\eta} \frac{4}{3}
      \frac{\alpha_S}{2\pi}
      \left(\frac{1 + \eta^2}{1 - \eta}\right)_+ f_i (\xi/\eta, Q^2) \\
  \frac{\partial D(\xi,Q^2)}{\partial\ln Q^2}
    & = & \int\limits_\xi^1 \frac{d\eta}{\eta} \frac{\alpha_{QED}}{2\pi}
      \left(\frac{1 + \eta^2}{1 - \eta}\right)_+ D (\xi/\eta, Q^2) \\
\end{eqnarray}
which are summing the leading logarithms (\ref{eq:LL}) and their QCD
counterparts, respectively.

The Monte Carlo implementation of (\ref{eq:x-sect}) can now be
decomposed into five parts:
\begin{itemize}
  \item{} Generation of $(x,y,z)$ triples according to the
    distribution (\ref{eq:x-sect}).
  \item{} Generation of photons corresponding to the chosen $(z,Q^2)$.
  \item{} Generation of final state photons.
  \item{} Generation of QCD parton showers corresponding to the chosen
    $(x',Q^{\prime2})$.
  \item{} Hadronization of the generated partons.
\end{itemize}
It is now obvious that these problems can be solved by almost
independent programs.  Because the hadronically {\em inclusive\/} cross
section is known, the first three points can be handled by \KRONOS/
without any intervention from \HERWIG/\footnote{There is one subtlety,
however:  when \HERWIG/ reconstructs the initial state parton shower
{}from the chosen $(x',Q^{\prime2})$ pair, it might reject it due to
mismatches between the $Q^2$ ordering in the evolution equations of
the employed parton distribution functions and the angular ordering in
\HERWIG/'s evolution.  This is taken care of in \KROWIG/ by rejecting
the event.}\label{evol-mismatch}.
Similarly, \HERWIG/ needs no information from \KRONOS/
besides the $(x',Q^{\prime2})$ pair to reconstruct an appropriate
parton shower. Furthermore, as already stated above, in the parton model
hadronization happens with unit probability, once the hard subprocess
has been generated.

{}From this discussion a simple physical picture emerges: the Monte
Carlo event generator for radiative corrections at the parton level
can be used unchanged, because it only depends on the hadronically
inclusive cross sections.  The algorithms implementing this step in
\KRONOS/ have been described in \cite{ADM+92a}.  On the other hand, the
QCD Monte Carlo can completely ignore the leptonic radiative
corrections, provided it uses the electron {\em after\/} initial state
radiation as input.  Phrased differently, the leptonic radiative
corrections affect the QCD Monte Carlo only through broadening the sharp
electron energy spectrum into an energy distribution.


\subsection{Programming issues}
\label{sec:programming}

In order to exploit the simple physical picture of initial state
radiation as an effective beam energy spread, several requirements
must be met.

First of all, the called QCD Monte Carlo event generator must be very
disciplined in its use of {\em global variables\/}.  Any assumption
based on a fixed beam energy will defeat a simple interface.
Typically such assumptions would be used at initialization time to set
up internal tables, whose contents will be invalid for radiative
events.

\begin{figure}[tbp]
  \begin{center}
    \begin{picture}(100, 80)
      \put(  0,  0){\framebox(100, 70)[t]{\KRONOS/ driver program and
                                          parameter management\strut}}
      \put(  5,  5){\framebox( 90, 55)[t]{\.{krowig()} ($\approx$
                                          \.{HWIGPR})\strut}}
      \put( 10, 10){\framebox(  5, 40){}}
      \put( 20, 10){\framebox( 40, 40)[t]{\.{HWEPRO()}\strut}}
      \put( 25, 15){\framebox( 30, 25)[t]{\.{HWHDIS()}\strut}}
      \put( 30, 20){\framebox( 20, 10){\.{kronos()}}}
      \put( 65, 10){\framebox(  5, 40){}}
      \put( 75, 10){\framebox(  5, 40){}}
      \put( 85, 10){\framebox(  5, 40){}}
    \end{picture}
    \caption{Top level structure of \KROWIG/.}
    \label{fig:structure}
  \end{center}
\end{figure}
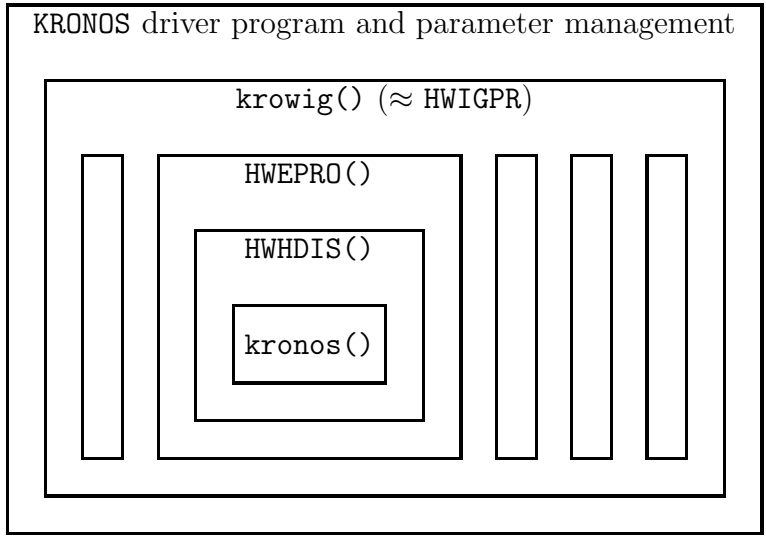

Fortunately, \HERWIG/ is a well-behaved program in these respects.
Since initial state radiation can only {\em reduce\/} the effective
beam energy, \HERWIG/'s table of Sudakov formfactors is {\em not\/}
invalidated by radiative events.  We can therefore use the following
simple multiply layered approach, which is shown graphically in
figure~\ref{fig:structure}.

The main routine is derived from the command interpreter and parameter
management routines of \KRONOS/, which have been described in
\cite{ADM+92a} in detail.  The only difference is the larger
number of parameters which are needed to control \HERWIG/\footnote{It
is important to note that this implies that the parameter common
blocks \.{/krpcom/} of \KRONOS/ and \KROWIG/ are {\em not\/}
compatible.} (cf.\ table~\ref{tab:herwig-params}).
Instead of calling the parton level \.{kronos()} event generation
routine, the main routine now calls the \.{krowig()} routine which
has been adapted from \HERWIG/'s main event generation loop. In the
innermost layer, \HERWIG/'s parton level generator is emulated by a
wrapper around the original \.{kronos()} parton level generator.

This simple structure is straightforwardly implemented because both
programs use the standard \hepevt/ event record \cite{AKV89}, which
allows a transparent flow of event information between Monte Carlo
components.  Otherwise more complicated translation routines would
have been necessary.


\section{Parameters}
\label{sec:parameters}


\subsection{\KRONOS/ Parameters}
\label{sec:kronos-parameters}

\begin{table}
  \begin{minipage}{\textwidth}
  \begin{center}
  \begin{tabular}{|c|c|c|}
    \hline\hline
    Variable name   & semantics             & Default value
    \\\hline\hline
    \.{ahpla}       & $1/\alpha_{QED}$      & 137.0359895
    \\\hline
    \.{lambd}       & $\Lambda_{QCD}$       & 0.18 GeV
    \\\hline
    \.{mass1e}      & $m_{e^\pm}$           &
                                         $0.51099906\cdot 10^{-3}$ GeV
    \\\hline
    \.{mass1p}      & $m_p$                 & 0.93827231 GeV
    \\\hline
    \.{mass1z}      & $M_{Z^0}$             & 91.17 GeV
    \\\hline
    \.{mass1w}      & $M_{W^\pm}$           & 80.0 GeV
    \\\hline
    \.{sin2w}       & $\sin^2\theta_W$      & 0.23 GeV
    \\\hline
        \.{elecen}      & $e^-$ energy          & 30 GeV
    \\\hline
        \.{proten}      & $p$ energy          & 820 GeV
    \\\hline
        \.{nevent}      & Number of events      & 1000
    \\\hline
        \.{bstyle}      & ``branching style''\footnote{%
        For reasons of backward compatibility, this parameter has
        slightly confusing semantics associated with it.  It should be
        noted however, that the value of \.{bstyle} is {\em not\/} at
        the discretion of the user.  The only supported values are
        \begin{itemize}
          \item [4] Full radiative corrections switched on (default).
          \item [0] No radiative corrections, Born cross section.
        \end{itemize}
        All other values will be reset to the default value.}  & $4$
    \\\hline
        \.{epsiln}      & Internal infrared cutoff & $10^{-4}$
    \\\hline
        \.{cutxmi}      & Lower cut in $x$      & 0.001
    \\\hline
        \.{cutxma}      & Upper cut in $x$      & 0.5
    \\\hline
        \.{cutymi}      & Lower cut in $y$      & 0.1
    \\\hline
        \.{cutyma}      & Upper cut in $y$      & 0.9
    \\\hline
        \.{cutqmi}      & Lower cut on hadronic $Q^2$     & 100 GeV$^2$
    \\\hline
        \.{cutqma}      & Upper cut on hadronic $Q^2$     & $s$
    \\\hline
        \.{rseed}       & Random number seed   & 54217137
    \\\hline
        \.{errmax}      & maximum error count  & 100
    \\\hline
        \.{verbos}      & verbosity            & 0
    \\\hline
        \.{runid}       & run identification       &
    \\\hline
        \.{stdin}       & standard input           & 5
    \\\hline
        \.{stdout}      & standard output          & 6
    \\\hline
        \.{stderr}      & standard error           & 6
    \\\hline
  \end{tabular}
  \end{center}
  \end{minipage}
  \caption{\KRONOS/ parameters}
\label{tab:kronos-parm}
\end{table}

A detailed discussion of the \KRONOS/ specific parameters can be found
in \cite{ADM+92a,ADM+92c} and will not be repeated here.
The parton distribution functions are taken from the {\tt PAKPDF}
library \cite{Cha92} and can be set with the usual \KRONOS/ commands,
which are also explained in detail in \cite{ADM+92a,ADM+92c}.
Note however, that the optional use of phenomenological
parametrizations of electroproduction {\em structure functions\/}
$F_i$ is not supported, because \KRONOS/ needs {\em parton
distributions\/} for generating the struck quark.


\subsection{\HERWIG/ Parameters}
\label{sec:herwig-parameters}

\begin{table}
  \begin{minipage}{\textwidth}
  \begin{tabular}{|c|c|c|c|}
    \hline\hline
    variable & \HERWIG/ & semantics & Default value
    \\\hline\hline
    \.{herwig}      & N/A
        & call \KROWIG/\footnote{Iff this variable is false, then
          \KROWIG/ behaves identical to \KRONOS/.} & true \\\hline
    \.{sue}         & N/A\footnote{This is implemented by adding
          10000 to the \HERWIG/ variable \.{IPROC} iff \.{sue}
          is false.}
        & generate S.U.E. & false \\\hline
    \.{elecen}      & \.{PBEAM1}
        & electron beam energy  & 30 GeV \\\hline
    \.{proten}      & \.{PBEAM2}
        & proton beam energy    & 820 GeV \\\hline
    \.{nevent}      & \.{MAXEV }
        & number of events  &  1000 \\\hline
    \.{lambd }      & \.{QCDLAM}
        & $\Lambda_{QCD}$  & 0.180 GeV \\\hline
    \.{hwvqcu}      & \.{VQCUT }
        & quark virtuality cutoff  & 0.48 GeV \\\hline
    \.{hwvgcu}      & \.{VGCUT }
        & gluon virtuality cutoff  & 0.10 GeV \\\hline
    \.{hwvpcu}      & \.{VPCUT }
        & photon virtuality cutoff & $10^{10}$
          GeV\footnote{i.e.~photons in {\em hadronic\/} final state
          parton showers are suppressed by default.} \\\hline
    \.{hwclma}      & \.{CLMAX }
        & maximum cluster mass  & 3.35 GeV    \\\hline
    \.{hwpspl}      & \.{PSPLT }
        & cluster mass distribution\footnote{See
          \cite{MWA+92} for a detailed description of this parameter.}
          & 1.00    \\\hline
    \.{hwqdiq}      & \.{QDIQK }
        & max.~scale for glue $\rightarrow$ diquark
          & 0.00 GeV\footnote{i.e.~no diquarks by default.} \\\hline
    \.{hwpdiq}      & \.{PDIQK }
        & glue $\rightarrow$ diquark rate & 5.00    \\\hline
    \.{hwqspa}      & \.{QSPAC }
        & spacelike evolution cutoff  & 2.50 GeV   \\\hline
    \.{hwptrm}      & \.{PTRMS }
        & intrinsic transverse momentum & 0.00 GeV \\\hline
    \.{hwenso}      & \.{ENSOF }
        & multiplicity of S.U.E.\footnote{See
          \cite{MWA+92} for a detailed description of this parameter}
          & 1.00    \\\hline
    \.{hwipri}      & \.{IPRINT}
        & \HERWIG/'s verbosity\footnote{See
          \cite{MWA+92} for a description of this parameter.
          The default value $0$ shuts up \HERWIG/ almost completely.}
          & 0    \\\hline
    \.{hwmaxp}      & \.{MAXPR }
        & \#\ of events printed  & 0    \\\hline
    \.{hwmaxe}      & \.{MAXER }
        & maximum \#\ of errors & 100    \\\hline
    \.{hwlwev}      & \.{LWEVT }
        & unit for writing output events\footnote{This option
          should {\em not\/} be used because radiated photons might
          be missing and \HERWIG/'s internal coordinate system is
          different from \KROWIG/'s (\KROWIG/ has the incoming proton
          in the $+z$ direction, whereas \HERWIG/ assumes that the
          incoming proton goes in the $-z$ direction).}
          & 0\footnote{i.e. don't write events to disk.}
            \\\hline
    \.{hwlrsu}      & \.{LRSUD }
        & unit for restoring formfactors & 0    \\\hline
    \.{hwlwsu}      & \.{LWSUD }
        & unit for saving formfactors & 77    \\\hline
    \.{hwazso}      & \.{AZSOFT}
        & soft gluon correlations & true    \\\hline
    \.{hwazsp}      & \.{AZSPIN}
        & gluon spin correlations & true    \\\hline
    \.{hwncol}      & \.{NCOLO }
        & \# of colors  & 3\footnote{You probably don't want to change
        this!} \\\hline
    \.{hwnfla}      & \.{NFLAV }
        & \# of active flavours  & 6    \\\hline
  \end{tabular}
  \end{minipage}
  \caption{\HERWIG/ parameters accessible through
           \.{krdcmd ('set ...')}}
\label{tab:herwig-params}
\end{table}

The variables in table \ref{tab:herwig-params} fall into three categories
\begin{itemize}
  \item {} \.{herwig} and \.{soe} have no analogue in \HERWIG/, they
    control whether and how \HERWIG/ is called to generate hadronic final
    states from \KRONOS/' partons.
  \item {} some \HERWIG/ variables have denote the same physical
    parameter as a \KRONOS/ variable; here the former is set from the
    latter.
  \item {} finally there are plenty of variables in \HERWIG/ that
    have no analogue in \KRONOS/: the name of these variables is
    derived from their name in \HERWIG/ by prepending \.{`hw'} and
    stripping the fifth and sixth character (this does not lead to
    ambiguities for the variables considered).
\end{itemize}

\HERWIG/ variables which do not affect DIS have not been made
available in the \.{krdcmd()} interface.

Note that the variable \.{VPCUT} relates to photons radiated from the
{\em hadronic\/} subsystem and does {\em not\/} affect the {\em
leptonic\/} side, which is handled by \KRONOS/.


\subsection{Tuning of parameters}
\label{sec:tuning}

In the Monte Carlo generator working group \cite{MBB+92} of the
1991 HERA workshop \HERWIG/ version 5.3 has been tuned to EMC data,
using the KMRS B0 parton distributions.  The
following two sets of changes with respect to the \HERWIG/ version 5.3
default values have been proposed:
\begin{enumerate}
   \item {} No soft underlying event:
     \begin{verbatim}
     # herawg10-set1.krowig
     set sue false
     set hwclma 2.00
     set hwqspa 2.00
     set hwpspl 0.50
     set hwptrm 0.70
     \end{verbatim}
     In addition, it has been proposed to set the parameter \.{BTCLM}
     in the subroutine \.{HWCCUT()} to 3.00, while increasing \.{CLMAX}
     to 3.00.  However \KROWIG/ does {\em not\/} support the
     modification of \.{BTCLM} because it
     is {\em not\/} a tunable \HERWIG/ parameter and changing it
     requires modification of the \HERWIG/ sources.  This should be
     done by the user at his own risk.  If
   \item {} With soft underlying event:
     \begin{verbatim}
     # herawg10-set2.krowig
     set sue true
     set hwclma 2.00
     set hwptrm 0.40
     set hwenso 0.60
     \end{verbatim}
     This set gives a slightly worse agreement with EMC data.
\end{enumerate}

It should be noted however, that the default parameter values have been
changed from \HERWIG/ 5.3 to \HERWIG/ 5.4.  Therefore a retuning might
be necessary.


\section{\f77/ Interface}
\label{sec:f77}

The application program interface of \KROWIG/ is almost identical to
\KRONOS/' \cite{ADM+92a}.  On the top level, where \KROWIG/ is
controlled by character strings passed to \.{krdcmd()} (see figure
\ref{ex:f77}), the interface is identical, except for the possibility
to change \HERWIG/ parameters.  These parameters are collected in
table \ref{tab:herwig-params}.  The {\tt generate} command will
execute a loop similar to the one shown in figure \ref{ex:f77-krdcmd}.
The generated event will be passed to the application program in the
standard \hepevt/ common block \cite{AKV89}.

\begin{example}{\f77/ interface}{ex:f77}
\.{* krowigappl.f}                                                \\
\>\>\> \.{...}                                                    \\
\>\>\> \.{call krdcmd ('set maxpr 1')}   \C instruct \HERWIG/ to
                                            dump the first event  \\
\>\>\> \.{call krdcmd ('init')}   \C initialize the generator     \\
\>\>\> \.{...}                                                    \\
\>\>\> \.{call krdcmd ('generate 10000')}
                                   \C generate 10000 events       \\
\>\>\> \.{...}                                                    \\
\>\>\> \.{call krdcmd ('close')}  \C cleanup                      \\
\>\>\> \.{...}
\end{example}

At the lower level function call interface there are also almost no
changes from \KRONOS/.  The call to the \.{kronos()} subroutine is to
be replaced by a call to the wrapper routine \.{krowig()} which
calls \.{kronos()} and the appropriate \HERWIG/ routines.  The
single integer parameter is interpreted as follows:
\begin{itemize}
  \item [0:] initialize the generator and write an initialization
      record to \hepevt/.
  \item [1:] generate an event and store it in \hepevt/.
  \item [2:] perform final calculations and write the results
      to \hepevt/.
\end{itemize}

\begin{example}{Event generation loop}{ex:f77-krdcmd}
\.{* krdcmd.f}                                                    \\
\>\>\> \.{subroutine krdcmd (cmdlin)}                             \\
\>\>\> \.{character*(*) cmdlin}                                   \\
\>\>\> \.{...}                                                    \\
\>\>\> \.{else if (cmdlin.eq.'gen')}                              \\
\>\>\> \>\> \.{do 10 n = 1, nevent}                               \\
\>\>\> \>\> \>\> \.{call krowig (1)}   \C generate an event       \\
\>\>\> \>\> \>\> \.{call hepawk ('scan')} \C analyze the event    \\
\.{10} \>\>\>\>\> \.{continue}                                    \\
\>\>\> \.{else}                                                   \\
\>\>\> \.{...}                                                    \\
\>\>\> \.{end}
\end{example}

As in any interface of independly developed programs, there remain
some rough edges.  While these rough edges could be avoided in areas
where accepted standards exist (e.g.~the passing of event information
via \hepevt/ \cite{AKV89} is by now well established), the control of
input parameters still leaves something to be desired: the fine
grained control over reinitializations in \KRONOS/ (see \cite{ADM+92a}
for a description) can not be achieved with \HERWIG/, unless we allow
to make some (simple) modifications to the \HERWIG/ sources.  Because
it is desirable to keep these untouched\footnote{This should allow
\KROWIG/ to work with future \HERWIG/ versions unmodified.}, we let
modifications of all \HERWIG/ parameter cause a reinitialization,
even if not strictly necessary.


\section{Limitations of \KROWIG/ \KWVersion/}
\label{sec:limitations}

\subsection{Compton channel}
\label{sec:no-compton}

\KROWIG/ is not applicable in the kinematical region corresponding to
the so-called Compton events:
\begin{equation}
  (p_e + p_\gamma)^2 \gg Q^2_{hadr.} \rightarrow 0
  \label{eq:compton-region}
\end{equation}
In this region the quark parton model is not applicable and therefore
\HERWIG/ does not provide a realistic model for the generation of
hadronic final states.  For studies of the leptonic and photonic final
states of these events it is recommended to use the stand-alone
\KRONOS/ with the structure function ({\em not\/} parton
distributions) option.

Future revisions of \KROWIG/ might overcome this restriction by
exploiting a simple model (maybe along the lines of the \HERWIG/
remnant fragmentation -- a.k.a.~``soft underlying event'').

\subsection{${\cal O}(\alpha_S)$ matrix elements}
\label{sec:no-alphas}

\KROWIG/ can {\em not\/} yet be used with the ${\cal O}(\alpha_S)$ matrix
element option of recent \HERWIG/ versions.  This is not a fundamental
limitation and could be overcome in future versions of \KROWIG/, after
the ${\cal O}(\alpha_S)$ QCD inclusive cross sections have been
implemented in \KRONOS/.

\subsection{Charged currents}
\label{sec:no-CC}

No charged current events will be generated.  This restriction will be
lifted (without any changes to \KROWIG/), once the charged current
subsystem of \KRONOS/ has been released.


\section{Conclusions}
\label{sec:concl}

We have presented version \KWVersion/ of the Monte Carlo event
generator \KROWIG/ for deep inelastic scattering at HERA energies.
\KROWIG/ acts as an interface for the QED generator \KRONOS/ and the
QCD generator \HERWIG/.  It is suitable for the study of hadronic
final states, taking into account the bulk of the electro-weak
radiative corrections.

\KROWIG/ is complementary to the {\tt DJANGO} Monte Carlo,
because it includes higher order QED corrections and uses a different
QCD Monte Carlo.


\section*{Availability}

The latest release of \KROWIG/ is available from the author upon
request.  In addition, \KROWIG/ is available by anonymous ftp from
{\tt freehep.scri.fsu.edu} in the directory {\tt
freehep/event\_generators/krowig}.  It is nevertheless recommended to
notify the author, in order to be informed of future bug fixes and
enhancements.


\appendix

\section{Revision History}
\label{sec:History}

\subsection*{Version 1.0, June 1992}

First official release.


\section{Installation}
\label{sec:Installation}

\KROWIG/ is distributed in \.{PATCHY} \cite{KZ88} CARDS format.  The
installion scripts (a {\tt Makefile} for UNIX, a {\tt BUILD.COM} DCL
file for VMS, and a JCL/370 deck for MVS) assume the availability of
the \KRONOS/ CARDS file (version 1.2 or later) and of the
\HERWIG/\footnote{Usually available from {\tt
CBHEP::DISK\$USER1:[THEORY.HERWIG.PATCHY]HERWIG54.CAR} via DECNET.}
(version 5.4) sources in CARDS format.  Furthermore
{\tt PAKPDF} \cite{Cha92} and \hepawk/ \cite{Ohl92a,Ohl92b} are expected
as linkable libraries.  The latter can be
replaced by any other event analysis subroutine.

If PATCHY is not available, a \f77/ version of the source can be
provided by the author.

\subsection{Building \HERWIG/}
\label{sec:inst-herwig}

In a first step, \HERWIG/ should be compiled {\em without\/} the
functions that will be overwritten by \KROWIG/: \.{hwhdis},
\.{pdfset}, and \.{structf}.  This can be done by removing them from
the \f77/ distribution or by using the CARDS file with the CRADLE
\begin{verbatim}
   +EXE.
   +USE,IBM,IBMMVS.
   +USE,*HERWIG.
   +USE,P=HERWIG,D=HWHDIS,T=INHIBIT.
   +USE,P=HERWIG,D=STRUCTF,T=INHIBIT.
   +USE,P=HERWIG,D=PDFSET,T=INHIBIT.
   +PAM,11,T=CARDS.
   +QUIT.
\end{verbatim}
If the standard \HERWIG/ distribution runs on the target system, no
changes should be necessary for using it in \KROWIG/.

\subsection{Building \KROWIG/}
\label{sec:inst-krowig}

After adjusting filenames in the installation scripts, these scripts
should build \KROWIG/ without further user intervention.
If \HERWIG/ is not available in CARDS format, the \HERWIG/ common
blocks have be imported manually into the routines requesting them
with \.{+CDE,HERCDES}.  Note that it is {\em not\/} possible to link
\KROWIG/ with the standard \KRONOS/ object module because the
parameter common block \.{/krpcom/} has to be enlarged and the call to
the \.{kronos()} subroutine has to be replaced by the \.{krowig()}
wrapper.


\section{Common Blocks and Subroutines}
\label{sec:ext-names}

To avoid possible name clashes with other packages, all external
symbols exported by \.{KRONOS} begin with the two letters \.{KR}.
This convention is also followed by \KROWIG/, except where routines
{}from other programs are overwritten.


\begin{itemize}
  \item Common Blocks: \hfil\goodbreak
    The following common block is used by \KROWIG/ in addition to
    \KRONOS/' and \HERWIG/'s common blocks.
    \begin{itemize}
      \item \.{/krcevt/}: small event record in the style of
        \hepevt/ for hiding the photons from \HERWIG/.
      \item \.{/krcwig/}: a collection of variables used for
        non-local communication between \KRONOS/, \HERWIG/,
        and \KROWIG/.
    \end{itemize}
  \item Primary entry point: \hfil\goodbreak
    \begin{itemize}
      \item \.{krowig}: main entry point, replacing \.{kronos} (which it
        calls itself).
    \end{itemize}
  \item Utility routines: \hfil\goodbreak
    \begin{itemize}
      \item \.{kr2hw1}: glue routine, editing \KRONOS/' output
        in \hepevt/ to match \HERWIG/'s expectations.
      \item \.{kr2hw2}: glue routine, fixing \hepevt/ after \HERWIG/
        has processed it.
      \item \.{kr2ent}: utility routine, used in \.{kr2hw1} for
        filling \hepevt/.
    \end{itemize}
  \item Overwritten entry point: \hfil\goodbreak
    \begin{itemize}
      \item \.{hwhdis}: overwritten standard \HERWIG/ deep inelastic
        subprocess.
      \item \.{hwaend}: overwritten \HERWIG/ user exit.
      \item \.{pdfset, structf}: emulation of {\tt PDFLIB} \cite{Plo91}
        routines (used in \HERWIG/) by {\tt PAKPDF} \cite{Cha92}.
    \end{itemize}
\end{itemize}


\section{Example}
\label{sec:sample}

As an example application of \KROWIG/, we demostrate how the generate
a rapidity distribution and the so-called ``seagull-plot''.

\subsection{{\tt sample.krowig}}

Here is a simple \KROWIG/ command file, setting up parameters and
generating 1000 events.

{\small
\begin{verbatim}
# sample.krowig -- Example KROWIG 1.0 command file
#
# nominal HERA energies
set elecen  30.0
set proten 820.0
#
# no. of events
set nevent 1000
#
# of course we want QCD included
set herwig true
set hwmaxe 100
set hwmaxp 0
set hwlrsu 0
set hwlwsu 0
#
# Morfin/Tung `B2'
struct mt 4
#
# kinematical cuts
set cutqmi 20.0
set cutymi  0.015
set cutyma  0.99
set cutxmi  0.7e-4
set cutxma  0.99
#
# go!
init
generate
#
# we're done.
exit
\end{verbatim}
} 

\subsection{{\tt sample.hepawk}}

This is a simple \hepawk/ script, selecting events that satisfy
semi-realistic acceptance cuts and filling histograms for charged
rapidity distributions and the famous ``seagull plot''\footnote{Note
that this \hepawk/ script requires version 1.2 or later of \hepawk/
\cite{Ohl92b}, because the Lorentz boost operator \.{/|} is not
available in version 1.0, which has been described in \cite{Ohl92a}.}.

{\small
\begin{verbatim}
# sample.hepawk -- sample HEPAWK analyzer for KROWIG v1.0

BEGIN
  {
    printf ("\nWelcome to the KROWIG test:\n");
    printf ("****************************\n\n");

    printf ("Monte Carlo Version: %s\n", REV);
    printf ("                Run: %d\n", RUN);
    printf ("               Date: %s\n\n", DATE);

    x_min = 0.0001;
    y_min = 0.02;
    Q2_min = 25.0;
    W2_min = 0.0;
    theta_min_had = theta_min_em = 6.0 / DEG;
    theta_max_had = theta_max_em = 174.0 / DEG;

    h_rap_cms = book1 (0, "rapidity (cms)", 50, -10, 10);
    h_sg = book1 (0, "Seagull", 50, -1, 1);
    h_xf = book1 (0, "xf", 50, -1, 1);

    incut = 0;               # initialize counter
  }


  {
    S = (@B1:p + @B2:p)^2;
    found_electron = 0;
    for (@e in ELECTRONS)    # Collect the outgoing electron.
      {
        $q = @B1:p - @e:p;
        Q2 = - $q^2;
        x = Q2/(2*$q*@B2:p);
        y = Q2/(x*S);
        if (Q2_min <= Q2 && x_min <= x && y_min <= y
            && theta_min_em <= angle (@e:p, @B2:p) <= theta_max_em)
          {
            found_electron = 1;
            break;
          }
      }

    if (!found_electron)
      next;
  }


  {
    # Step 1: find hadronic center of mass system:

    $p_hadr = $NULL;         # Collect (charged) hadronic momentum
    for (@p in HADRONS)
      if (charge (@p:id))
        if (theta_min_had <= angle (@p:p, @B2:p) <= theta_max_had)
          $p_hadr += @p:p;

    W2 = $p_hadr^2;
    if (W2 < W2_min)
      next;
    incut++;  # We've passed all inclusice cuts, count this event!

    if ($p_hadr:0 == 0)      # no hadrons in the cuts,
      next;                  # skip this event
    W = sqrt (W2);

    # Step 2: boost to hadronic CMS

    $beta = $p_hadr / $p_hadr:0;
    $q_phot = ($p_hadr - @B2:p) /| $beta;   # virt. photon in hadr. CMS
    $q_phot -= $E0 * $q_phot:0;             #   - make it a 3-vector
    $q_phot /= sqrt (- $q_phot^2);          #   - and normalize

    for (@p in HADRONS)
      if (charge (@p:id))
        if (theta_min_had <= angle (@p:p, @B2:p) <= theta_max_had)
          {
            $p = @p:p /| $beta;             # momentum in hadr. CMS
            fill (h_rap_cms, rap ($p));

            xf = - 2 * $p * $q_phot / W;  # project on virt. photon
            # extract transversal component and square it
            $p:0 = 0;
            pt2 = - ($p + $q_phot * ($q_phot * $p))^2;
            fill (h_sg, xf, 0, pt2);
            fill (h_xf, xf);
          }
  }

END
  {
    # Dump some numbers
    printf ("\nRESULTS:\n");
    printf ("********\n\n");
    printf ("       raw events: %d, raw cross section:        %g mb\n",
            NEVENT, XSECT);
    printf ("            error: %g%%\n", (ERROR/XSECT) * 100.0);
    printf ("events after cuts: %d, cross section after cuts: %g mb\n",
            incut, (incut/NEVENT) * XSECT);
    printf ("\nHISTOGRAMS:\n");
    printf ("***********\n\n");

    scale (1/NEVENT);

    # finish the seagull plot:
    arith (h_sg, h_sg, "/", h_xf);
    delete (h_xf);

    plot ();   # plot the histograms

    printf ("\ndone.\n");
  }
\end{verbatim}
} 

\subsection{{\tt sample.output}}

The following output should result from the input files above, modulo
small roundoff errors.  Note that the warning messages from \HERWIG/
correspond to the mismatches in the QCD evolution mentioned in the
footnote on page \pageref{evol-mismatch} and are harmless.

{\small
\begin{verbatim}
krdcmd: message: starting KRONOS, Version  1.02/00, (build 920626/0055)
hepawk: message: starting HEPAWK, Version  1.02/00, (build 920623/0117)
krowig: message: starting KROWIG, Version  0.99/00, (build 920626/0055)

Welcome to the KROWIG test:
****************************

Monte Carlo Version: v00.99 (Jun 26 00:00:00  1992)
                Run: 1035996352
               Date: Jun 26 00:56:00  1992



          HERWIG 5.4    JANUARY   1992

          PARTICLE TYPE  21=PI0  SET STABLE

          INPUT EVT WEIGHT   =   .1000E+01
          INPUT MAX WEIGHT   =   .1000E+01

 HWWARN CALLED FROM SUBPROGRAM HWSBRN: CODE = 105
 EVENT      95:   SEEDS =  382793286 & 1741717887  WEIGHT =  .1000E+01
 EVENT KILLED.   EXECUTION CONTINUES

 HWWARN CALLED FROM SUBPROGRAM HWSBRN: CODE = 101
 EVENT     228:   SEEDS = 1689384423 & 1215685145  WEIGHT =  .1000E+01
 EVENT KILLED.   EXECUTION CONTINUES

 HWWARN CALLED FROM SUBPROGRAM HWBGEN: CODE = 100
 EVENT     310:   SEEDS =  929482140 &  326931761  WEIGHT =  .1000E+01
 EVENT KILLED.   EXECUTION CONTINUES

 HWWARN CALLED FROM SUBPROGRAM HWSBRN: CODE = 105
 EVENT     360:   SEEDS =  255018931 & 1779155573  WEIGHT =  .1000E+01
 EVENT KILLED.   EXECUTION CONTINUES

 HWWARN CALLED FROM SUBPROGRAM HWSBRN: CODE = 105
 EVENT     381:   SEEDS =  915725380 &  739707742  WEIGHT =  .1000E+01
 EVENT KILLED.   EXECUTION CONTINUES

 HWWARN CALLED FROM SUBPROGRAM HWSBRN: CODE = 105
 EVENT     519:   SEEDS =  461210881 & 1073328688  WEIGHT =  .1000E+01
 EVENT KILLED.   EXECUTION CONTINUES

 HWWARN CALLED FROM SUBPROGRAM HWBGEN: CODE = 100
 EVENT     690:   SEEDS =  281115117 & 1525150099  WEIGHT =  .1000E+01
 EVENT KILLED.   EXECUTION CONTINUES

 HWWARN CALLED FROM SUBPROGRAM HWSBRN: CODE = 105
 EVENT     805:   SEEDS =  485884270 & 1401730461  WEIGHT =  .1000E+01
 EVENT KILLED.   EXECUTION CONTINUES

\end{verbatim}
} 
\newpage
{\scriptsize
\baselineskip=0.9\baselineskip
\begin{verbatim}
RESULTS:
********

       raw events:       1000, raw cross section:         .4587E-04 mb
            error:  .4994    %
events after cuts:        720, cross section after cuts:  .3303E-04 mb

HISTOGRAMS:
***********

rapidity (cms)

 HBOOK     ID =         1                                        DATE

        8.4                              9
        8.2                              X6
        8                               8XX
        7.8                             XXX
        7.6                             XXX
        7.4                             XXX
        7.2                            9XXX
        7                              XXXX5
        6.8                            XXXXX
        6.6                            XXXXX
        6.4                            XXXXX
        6.2                            XXXXX
        6                              XXXXX
        5.8                           7XXXXX6
        5.6                           XXXXXXX
        5.4                           XXXXXXX
        5.2                           XXXXXXX
        5                             XXXXXXX
        4.8                           XXXXXXX
        4.6                          9XXXXXXX
        4.4                          XXXXXXXX
        4.2                          XXXXXXXX0
        4                            XXXXXXXXX
        3.8                          XXXXXXXXX
        3.6                          XXXXXXXXX
        3.4                         9XXXXXXXXX
        3.2                         XXXXXXXXXX
        3                           XXXXXXXXXX
        2.8                         XXXXXXXXXX
        2.6                         XXXXXXXXXX
        2.4                         XXXXXXXXXX
        2.2                         XXXXXXXXXX
        2                           XXXXXXXXXX
        1.8                        0XXXXXXXXXX4
        1.6                        XXXXXXXXXXXX
        1.4                        XXXXXXXXXXXX
        1.2                        XXXXXXXXXXXX
        1                          XXXXXXXXXXXX
         .8                       6XXXXXXXXXXXX3
         .6                       XXXXXXXXXXXXXX
         .4                      6XXXXXXXXXXXXXX
         .2                    15XXXXXXXXXXXXXXX920

 CHANNELS  10   0        1         2         3         4         5
            1   12345678901234567890123456789012345678901234567890

 CONTENTS   1.                     134577886541
 *10**  1   0   00000000000000001376357193197066100000000000000000
            0   00000000000000031331995979302197941000000000000000

 LOW-EDGE       --------------------------
           10   1
            1.  09988877666554443322211     1122233444556667788899
            0   06284062840628406284062840482604826048260482604826

 * ENTRIES =      13490      * ALL CHANNELS =  .6745E+01      * UNDERFL
 * BIN WID =  .4000E+00      * MEAN VALUE   =  .5607E-01      * R . M .
\end{verbatim}
} 
\newpage
{\scriptsize
\baselineskip=0.9\baselineskip
\begin{verbatim}
Seagull

 HBOOK     ID =         2                                        DATE

        4.6                                                 6
        4.5                                                 X
        4.4                                                 X
        4.3                                                 X
        4.2                                                 X
        4.1                                                 X
        4                                                   X
        3.9                                                 X
        3.8                                                 X
        3.7                                                 X
        3.6                                                 X
        3.5                                                 X
        3.4                                                 X
        3.3                                                 X
        3.2                                                 X
        3.1                                                 X 0
        3                                                   X X
        2.9                                                 X X
        2.8                                                 X X
        2.7                                                 X X
        2.6                                                 X X
        2.5                                                 X X
        2.4                                                 X X
        2.3                                                 X X
        2.2                                                 X X
        2.1                                                 X X
        2                                                   X X
        1.9                                                 X X
        1.8                                                1X X
        1.7                                                XX X
        1.6                                                XX X 2
        1.5                                                XX X X
        1.4                                           0  29XX X X
        1.3           1  7 2                          X  XXXX X X
        1.2         0 X  X X  3                       X 1XXXX X X
        1.1         X X  X X  X8                      X8XXXXX X X
        1           X X 4X X  XX                     9XXXXXXX X X
         .9         X X XX X  XX 5                 7 XXXXXXXX X X
         .8        9X X XX X55XX X               2 X8XXXXXXXX X X
         .7       6XX X XX XXXXX X71             X XXXXXXXXXX1X X
         .6       XXX X XX3XXXXX XXX             X9XXXXXXXXXXXX X
         .5      8XXX X5XXXXXXXX9XXX6         049XXXXXXXXXXXXXX X
         .4      XXXX0XXXXXXXXXXXXXXX7      50XXXXXXXXXXXXXXXXX X
         .3     2XXXXXXXXXXXXXXXXXXXXX7    3XXXXXXXXXXXXXXXXXXX X
         .2     XXXXXXXXXXXXXXXXXXXXXXX8329XXXXXXXXXXXXXXXXXXXX1X
         .1     XXXXXXXXXXXXXXXXXXXXXXXXXXXXXXXXXXXXXXXXXXXXXXXXX

 CHANNELS  10   0        1         2         3         4         5
            1   12345678901234567890123456789012345678901234567890

 CONTENTS   1.      1 1  1 1  11                      1111114 3 1
            0   24671324925277104866432111123344475879301337560150
            0   28690015473255389571677832935004929789081291610120
            0   70652615438725664302854714164789679110584093597460
            0   66856629942783135483331165078129549753798806027850

 LOW-EDGE       -------------------------
            1.  1
            0   09988877666554443322211000001122233444556667788899
            0   06284062840628406284062840482604826048260482604826

 * ENTRIES =      26980      * ALL CHANNELS =  .4134E+02      * UNDERFL
 * BIN WID =  .4000E-01      * MEAN VALUE   =  .1525E+00      * R . M .

done.
krdriv: message: bye.
\end{verbatim}
} 


\end{document}